# A Cooperative Spectrum Sensing Scheme Based on Compressive Sensing for Cognitive Radio Networks

Fatima Salahdine, Elias Ghribi, Naima Kaabouch
School of Electrical Engineering and Computer Science
University of North Dakota, Grand Forks, United States
fatima.salahdine@und.edu, elias.ghribi@und.edu, naima.kaabouch@und.edu

**ABSTRACT**

In this paper, a cooperative spectrum sensing scheme based on compressive sensing is proposed. In this scheme, secondary users (SUs) are organized in clusters. In each cluster, SUs forward their compressed signals to the cluster head. Then, the cluster heads report the local sensing data to the fusion center to make the final decision. Analog to information converter is adopted to acquire the received SUs signals. The proposed scheme is evaluated using several metrics, including probability of detection, compression ratio, probability of error, and processing time. The simulation results are compared with those of other cooperative sensing schemes. With our proposed scheme, cooperative spectrum sensing achieves a high probability of detection of 97%, a processing time of 2ms, a low compression ratio of 10%, and low error rate of 35 %. These results show that our proposed method outperforms the spectrum sensing without compressive sensing.

## 1 INTRODUCTION

Cognitive radio will play an important role in the fifth generation (5G) communication systems. It enhances the access to the radio spectrum by allowing dynamic and opportunistic access to the spectrum by sensing and detecting free frequency channels [1], [2]. Spectrum sensing allows unlicensed users, secondary users (SUs), to sense the radio spectrum and identify channels not being used by their owners, primary users (PUs), at a specific period of time and at a specific location [3]. A number of spectrum sensing techniques have been proposed to detect the spectral opportunities within a wide frequency band of interest. Examples of these techniques include energy detection, matched filter, autocorrelation, and Euclidian distance-based sensing [1-5]. When selecting a sensing technique, a tradeoff should exist between complexity, accuracy, detection rate, false alarm rate, number of samples, and PU signal knowledge.

The sensing performance is impacted by several factors, including noise, uncertainty, spatial diversity, and multipath fading/shadowing. Detecting weak and noisy PU signals requires high SU receiver sensitivity, which can be achieved only by complex and costly hardware implementation [6]. To overcome these problems, cooperative spectrum sensing has been proposed as a potential solution to address the sensing inefficiency of each SU. It allows SUs to share their sensing measurements and take a common decision considering the interests of all the involved SUs. It aims to enhance the sensing efficiency and decision accuracy by exploiting the communication links between different SUs observing the same frequency band and belonging to the same geographic area [7], [8].

Over the last decade, a number of cooperative spectrum sensing schemes have been proposed in the literature. For instance, the authors of [9] proposed a distributed cooperative spectrum sensing where each SU performs the sensing and exchanges the results with the other SUs. However, this approach requires a great deal of processing time to converge toward a final decision. In [10], the authors proposed a centralized cooperative spectrum sensing scheme that requires a fusion center capable of sending and receiving data to and from SUs. SUs forward their sensing measurements to the fusion center where the channel state is determined, and the final decision is taken. This approach requires a fusion center with high computational capabilities. In [11], the authors proposed a relay assisted based scheme where relays are used to handle weak sensing and weak channel reporting between SUs. Some SUs also play the role of relays to receive and forward the received measurements to reach a distant SU with weak channel. This approach represents additional delays and requires more time to process by using intermediate relays. In addition, it can cause loss of data, communication overhead, and interference to the PUs.

In [12], the authors proposed a cooperative spectrum sensing technique based dual threshold in order to overcome the threshold mismatch limitation. The credibility metric is used to evaluate the proposed technique, which refers to the SUs with reliable sensing results. This technique adopts a dynamic dual threshold to minimize optimize the probability of error and find the optimum threshold, which may slow down the local spectrum sensing process and delay the cooperative sensing decision making. In [13], the authors proposed to decrease the cooperative spectrum sensing time in order to decrease the communication overload in the presence of hiding terminal problem. They used the human behavior based particle swarm optimization algorithm to minimize the probability of collision between different SUs. This technique converges to the optimal solution that requires less sensing time. However, it is costly, complex, and search time may be longer.

Compared to non-cooperative spectrum sensing, these cooperative techniques improve the sensing efficiency in terms of the detection rates, but represent higher processing time, larger data exchange, extra energy consumption, and system overhead. In addition, some of these techniques require advanced fusion centers able to process high computational signals. Moreover, most of these approaches sense only narrow bands, which limits their utilization.

Wideband spectrum sensing consists of sensing available channels within several frequency bands. To perform this wideband sensing, compressive sensing has been proposed as a low-cost sampling technique ensuring fast and efficient wideband sensing [14], [15]. It allows the extraction of the main information from a high dimensional sparse signal by removing the unnecessary information and reducing the sampling rate to a level below the Nyquist rate. The original signal can be then reconstructed at the receiver using a recovery algorithm [16]. A few cooperative spectrum sensing schemes have been proposed [17] in which compressive sensing was performed by each SU receiver to acquire only the main information from the received signal. In [6], the authors proposed a distributed cooperative sub-Nyquist spectrum sensing approach to deal with selective fading channels. This approach is based on a multi-hop technique within a large network to determine the diversity gain at each node. Its accuracy increases with the increase of the number of cooperating SUs. However, it represents additional system overhead and high processing time to perform the communications between all the cooperating SUs in a distributed architecture.

In [18], the authors proposed a centralized cooperative spectrum sensing where the autocorrelation function results of the SUs are sent to a fusion center for decision making. At this fusion center, the SUs' signals are recovered using sampling, Matching Pursuit recovery algorithm, and the sensing process is performed on the recovered signals. This technique is less sensitive to noise and adopts a simple recovery process. However, it requires high number of samples to compute the autocorrelation matrix, which complicates the sampling process and results in an additional processing time [19]. The authors in [20] proposed a spectrum sensing detection technique based on multi-rates to enable asynchronous sampling at each node. The compressed signals are forwarded to the fusion center to compute the test statistic and to perform the sensing process as well as the decisions making process. This technique represents low complexity with high detection rates. However, it requires advanced fusion centers with specific capabilities to store high number of measurements, perform spectrum sensing algorithms, and take final decisions.

The authors in [11] adopted a cooperative sensing technique with low complexity in recovering signals to achieve high cooperative gain and enhanced sensing results under Rayleigh fading channels. However, this approach requires the maximum number of the available channels to be known, which is not practically possible. In [21], a collaborative compressive sensing scheme was proposed to enhance the energy efficiency and energy consumption. It performs by finding the optimal parameters that maximize the energy saving. The energy efficiency is improved by reducing the number of samples due to compression; but, this technique requires high number of cooperative nodes and optimal values of compression ratios to operate.

These existing techniques give good results, can mitigate uncertainty due to multipath fading and shadowing, and can achieve satisfactory detection rates with low complexity in signal recovery. However, they represent high complexity in sampling, high processing time, and high communication overhead. Thus, there is a great need for more efficient techniques that perform the spectrum sensing in short delays and with higher detection rates.

In this paper, a cooperative wideband spectrum sensing scheme is proposed where the sensing is performed cooperatively and compressively in a centralized based scheme. This scheme is based on a clustering method to enhance the detection rates. It aims to address the tradeoff between the sensing time and the detection performance as well as the decision accuracy. The remainder of the paper is organized as follows. Section II reviews the theoretical background about the clustering based cooperative spectrum sensing and compressive sensing. Section III represents the methodology as well as the system model. In Section VI, the simulation results are discussed and compared based on several metrics. Finally, a conclusion is given at the end.

## 2 THEORETICAL BACKGROUND

### 2.1 Cooperative Spectrum Sensing

Each SU performs the local spectrum sensing independently of the other cooperating SUs in the network. The received signal by the i[th] SU can be expressed as

$$\begin{cases} x_i = h_i * s + n_i, H1 \\ x_i = n_i, H0 \end{cases} \quad (1)$$

where $x_i$ denotes the received signal by the i[th] SU, $h_i$ denotes the channel gain between the PU and the i[th] SU, s denotes the PU signal, n denotes the noise, and H1 and H0 denote the binary hypotheses related to the presence and absence of the PU signal, respectively.

Cooperative spectrum sensing has been proposed to overcome the negative effects of multi-path fading and improve the detection accuracy and efficiency [22], [23]. It allows several SUs to coordinate and exchange their spectrum sensing results in order to mitigate the fading uncertainty and improve the detection decision accuracy. Based on the spectrum sensing strategy, cooperative spectrum sensing schemes can be classified into three classes: (i) centralized; (ii) distributed; and (iii) external [24].

Centralized cooperative spectrum sensing scheme involves a fusion center to perform the final spectrum sensing and coordinate between the cooperating SUs. It can be performed through two strategies. In the first strategy, the cooperating SUs operate individually the local spectrum sensing to get the sensing results. Then, they report their local results to the fusion center to take the final decision through a reporting channel. The fusion center combines the received measurements to decide about the presence of the PU signal in the sensed channel [25]. A number of fusion rules can be considered such as AND rule, OR rule, and majority rule. The fusion center forwards the final decision to all the SUs. In the second strategy, the cooperating SUs directly send their received signals to the fusion center to perform the spectrum sensing. This strategy requires smart and advanced fusion centers able to perform such functions.

In the distributed cooperative spectrum sensing scheme, the cooperating SUs perform the local spectrum sensing and exchange their local sensing results in order to converge to a common final decision. External cooperative spectrum sensing scheme involves an external device to perform the spectrum sensing and take the decision. This device, also called external agent, forwards the detection decision to all the cooperating SUs, which limits the role of the SUs during the detection process.

The performance evaluation metrics used for cooperative spectrum sensing are derived from the performance evaluation metrics for individual spectrum sensing. These metrics are the probability of detection, the probability of false alarm, and the probability of miss detection. The individual probability of detection, $P_d$, can be expressed as the probability that an SU declares the presence of the PU signal when the spectrum is occupied. It is given as

$$P_d = \text{Prob}(H1/H1) \qquad (2)$$

where H1 denotes the hypothesis that the PU is present. The individual probability of false alarm, $P_{fa}$, can be expressed as the probability that the SU declares the presence of the PU signal when the spectrum is free. It is given as

$$P_{fa} = \text{Prob}(H1/H0) \qquad (3)$$

where H0 denotes the hypothesis that the PU is absent. The individual probability of miss detection, $P_{md}$, can be expressed as the probability that the SU declares the absence of the PU signal when the spectrum is occupied. It is given as

$$P_{md} = \text{Prob}(H0/H1) \qquad (4)$$

The cooperative probability of detection, $C_d$, is expressed as

$$C_d = 1-(1-P_d)^L \qquad (5)$$

where $P_d$ denotes the individual probability of detection and L denotes the number of cooperating SUs. The cooperative probability of false alarm, $C_{fa}$, is expressed as

$$C_{fa} = 1-(1-P_{fa})^L \qquad (6)$$

where $P_{fa}$ denotes the individual probability of false alarm and L denotes the number of cooperating SUs. The mathematical equation for each metric depends on which spectrum sensing technique is used.

## 2.2 Cooperative Spectrum Sensing Based Clustering

Cooperative spectrum sensing schemes represent satisfactory results in terms of detection rates. However, they perform well with a limited number of SUs. When the number of cooperating SUs is very high, these schemes become more complex with high processing time, which limits their efficiency. Thus, a cooperative spectrum sensing based clustering solution was proposed in [26-32]. SUs are combined in clusters based on some features such as the geographic area and the distance to the PU. For each cluster, one of the SUs plays the role of the cluster head (CH) in order to coordinate between its cluster's users and the fusion center. At each cluster, the SUs perform their spectrum sensing independently and then communicate their results to the cluster head. The cluster heads forward the results to the fusion center for final decisions [28]. Figure 1 represents an example of a cluster based cooperative spectrum sensing with 6 SUs divided into 3 clusters.

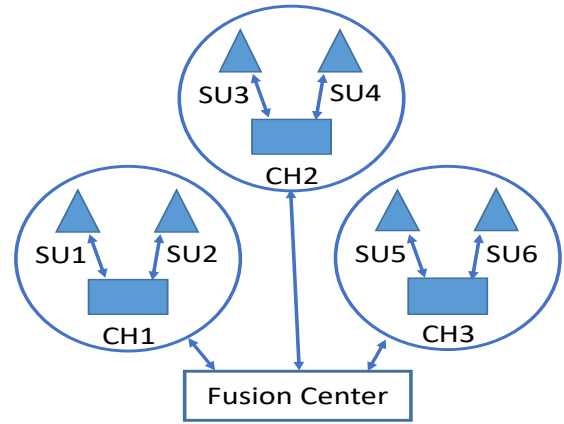

**Figure 1.** Example of cluster based cooperative spectrum sensing [26].

The performance evaluation metrics used for cooperative spectrum sensing based clustering are also derived from the performance evaluation metrics for individual spectrum sensing [30]. The cooperative probability of detection based clustering of the $k^{th}$ cluster, $Q_{d,k}$, is expressed as

$$Q_{d,k} = \int_0^\infty P_d(\gamma_k) f(\gamma_k) d\gamma_k \qquad (7)$$

where $P_d$ denotes the individual probability of detection, $\gamma_k$ denotes the SNR of the SU with the highest received signal power from the PU in the $k^{th}$ cluster, and $f(.)$ denotes the probability density function of $\gamma_k$, which can be expressed as

$$f(\gamma_k) = f(\max\{\gamma_{1,k}, \gamma_{2,k}, \dots, \gamma_{N_k,k}\}) \qquad (8)$$

where $N_k$ denotes the number of cooperating SUs in the $k^{th}$ cluster. Equation (7) can be reformulated as

$$f(k) = \frac{N_k}{\gamma} e^{(\frac{-\gamma_k}{\gamma})} [1 - e^{(\frac{-\gamma_k}{\gamma})}]^{N_k-1} \qquad (9)$$

where $\gamma$ denotes the average SNR. Moreover, the cooperative probability of detection based clustering of the $k^{th}$ cluster, $Q_{fa,k}$, is expressed as

$$Q_{fa,k} = \int_0^\infty P_{fa}(\gamma_k) f(\gamma_k) d\gamma_k \qquad (10)$$

where $P_{fa}$ denotes the individual probability of detection, $\gamma_k$ denotes the $k^{th}$ decision, f denotes, and j indicates the cluster [31]. The global probability of detection, $Q_d$, while taking into account all the clusters, can be derived from the previous equations as

$$Q_d = 1 - \prod_{k=1}^{J} 1 - Q_{d,k} \qquad (11)$$

where J is the number of the clusters and $Q_{d,k}$ denotes the probability of detection of the $k^{th}$ cluster. The global probability of false alarm, $Q_{fa}$, while taking into account all the clusters can be derived from the previous equations as

$$Q_{fa} = 1 - \prod_{k=1}^{J} 1 - Q_{fa,k} \qquad (12)$$

where J is the number of the clusters and $Q_{fa,k}$ denotes the probability of false alarm of the $k^{th}$ cluster.

## 2.3 Compressive Sensing

Compressive sensing approach has been proposed as a low cost solution for efficient and fast dynamic wideband spectrum sensing. It allows speeding up the scanning of the channels, reducing the computational complexity, and minimizing the hardware cost. This acquisition framework performs by acquiring only the main information from high dimensional sparse signals and neglecting the rest. Compressive sensing involves three main processes: sparse representation, signal sampling, and signal recovery. Compressive sensing requires sparse signals to perform. A sparse signal is a signal with more zero elements than non-zero elements. The signal acquisition is performed using sampling matrix techniques to extract the main information from the high dimensional signal. Compressed signal with few measurements can be then recovered at the receiver using recovery algorithms. Figure 2 illustrates the processes involved in compressive sensing.

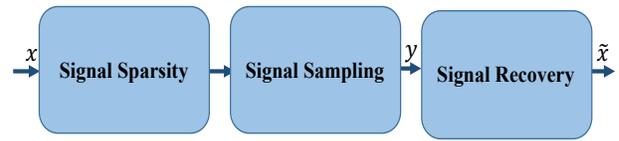

Figure 2. The compressive sensing processes [11].

Let consider a high dimensional sparse signal, x, of N samples. Through compressive sensing, the original signal, x, is acquired using a sampling matrix, $\varphi$, of MxN elements in order to extract M coefficients from N where M<<N. The compressed signal, also called the signal measurements, of M coefficients can be expressed as

$$y = \varphi x + n \qquad (13)$$

where y is the signal measurements, x is the sparse signal, $\varphi$ is the sampling matrix, and n denotes the noise. Figure 3 represents the signal sampling model.

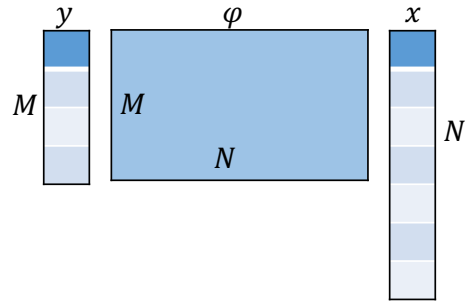

Figure 3. Signal sampling model.

A number of sampling matrices have been proposed, including random matrix [32], Circulant matrix [33], and Toeplitz matrix [14]. Random matrices are simple to design but they require high processing time and high memory capacity to store the random measurements. Circulant and Toeplitz matrices are structured matrices with determined form. They require less processing time and less measurements to sample the received signals.

The original signal can be then recovered from few measurements using the signal measurements. The recovery process performs by solving an underdetermined system and found its sparest solution. The underdetermined system is expressed as

$$\underset{x}{\text{minimize}} \ \|x\|_0 \ \text{subject to} \ y = \varphi x + n \quad (14)$$

where x denotes the original sparse signal of N coefficients, $\varphi$ denotes the sampling matrix of MxN elements, y is the signal measurements of M coefficients, and $\|.\|_0$ is the $l_0$-norm. Due to its NP hard nature, the underdetermined system can be solved by considering its convex relaxation using $l_1$-norm instead of $l_0$-norm. The underdetermined system ca be reformulated as

$$\underset{x}{\text{minimize}} \ \|x\|_1 \ \text{subject to} \ y = \varphi x + n \quad (15)$$

where $\|.\|_1$ is the $l_1$-norm and $\|.\|_p = \sqrt[p]{\sum |.|^p}$.

In order to estimate the original signal from few measurements and solve the underdetermined system, a number of recovery algorithms have been proposed. These algorithms can be divided into three classes: convex relaxation [25], Greedy [34], and Bayesian based recovery [35]. Convex relaxation algorithms are based on linear programming to solve the underdetermined system. They are suitable for any sampling matrix. They are accurate and consistent in terms of signal estimation. However, they require high processing time and computational complexity. Greedy algorithms consist of finding the local optimal and their corresponding position of non-zero coefficients. They require few measurements and less processing time to operate. However, they are not accurate. Bayesian based recovery algorithms consist of using Bayesian networks to estimate the unknown parameters using the available evidence and the linear relationship between these parameters. They are fast and accurate, but they require a prior knowledge of the relationship between the unknown parameters to estimate as well as the involved variables.

## 3 METHODOLOGY

Figure 4 presents the block diagram of the proposed cooperative spectrum sensing technique. Four main processes are involved: 1) signal acquisition with compressive sensing; 2) signal recovery and spectrum detection; 3) deciding and fusion center reporting; and 4) final decision making.

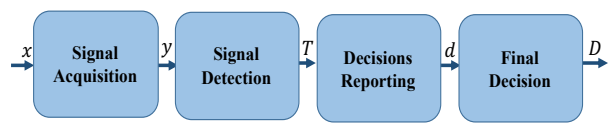

**Figure 4.** Cooperative spectrum sensing block diagram.

The proposed system consists of acquiring the SU received signal compressively by using an analog to information converter (AIC). Each SU sends its compressed signal to the cluster head, which is responsible of recovering the signal and performing the spectrum sensing from the SUs' measurements. Then, the cluster heads forward their local decisions simultaneously to a fusion center for the final decision about the spectrum occupancy. The general system model of the cooperative spectrum sensing is illustrated in Figure 5.

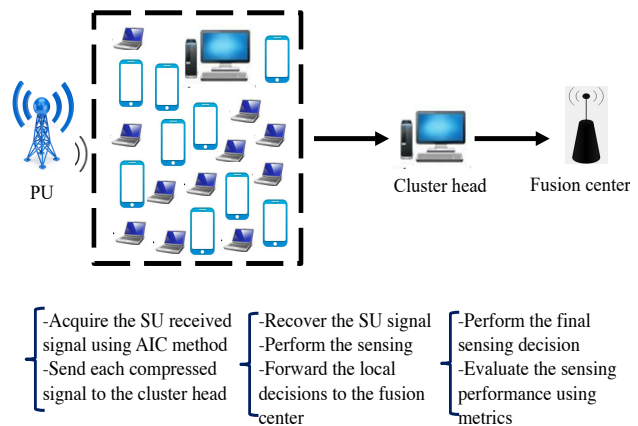

**Figure 5.** General model of the cooperative spectrum sensing based AIC technique.

The cluster heads receive the compressed signals from all the cooperating SUs in order to perform the spectrum sensing over all the SUs signals simultaneously. The cluster heads allow minimizing the processing time required for each SU to perform the spectrum detection and then each SU forwards its results to the fusion center. The cluster heads organize the measurements transmission as well as the data collection in order to report them to the fusion center. The fusion center, then receives these sensing measurements and performs the sensing decision of all the cooperating SUs. Therefore, our proposed approach will present less processing time, less data exchange, and less system overhead. Then, the necessity of using the cluster heads is justified. In addition, the compressive sensing was performed on all the cooperating SUs via the analog to information converter in order to perform the spectrum detection on the main information from the SU received signals instead of the local quantization method used in [36]. This local quantization method performed a generalized multi-bit cooperative sensing with local fusion center.

Let us consider a number of SUs, L, located at the same geographic area and sensing simultaneously the frequency band of interest. We assume this selected frequency band is owned and used by a PU. We opt for centralized cooperative spectrum sensing with no direct communication link between the cooperating SUs to report or exchange measurements or decisions. Cluster heads communicate only with the fusion center as the control unit of the sensing cooperation.

### 3.1 SUs Received Signals

Several SUs are aiming at sensing a selected frequency band. We assume there is no communication link between the SUs, which implies that the channels corresponding to different SUs are independent. The $i^{th}$ SU received signal, $y_i$, is expressed as

$$y_i = h_{pi}x_p + n_i \quad (16)$$

where $x_p$ denotes the PU signal, $h_{pi}$ denotes the complex channel gain of the sensing channel between PU and $i^{th}$ SU, and $n_i$ denotes the channel noise as illustrated in Figure 6.

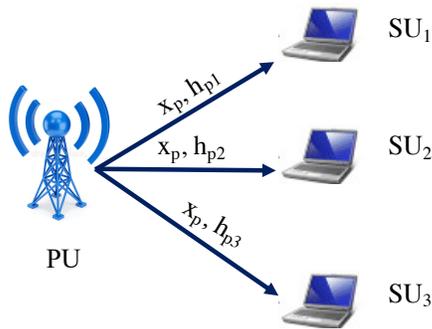

**Figure 6.** Example of cooperative system with a PU and three independent SUs.

Afterward, each SU transmits its signal to the cluster head. The received signal at the cluster head includes the PU signal, SUs signals, and noise. This signal, $Y_i$, can be expressed as

$$Y_i = h_{pi}x_p + n_i + \sum_{j=1, j \neq i}^{k} h_{ji}x_j + n_j \quad (13)$$

where $x_p$ denotes the PU signal, $x_j$ denotes the $j^{th}$ SU signal, $h_{pi}$ denotes the complex channel gain of the sensing channel between the PU and the $i^{th}$ SU, $h_{ji}$ denotes the complex channel gain of the sensing channel between the $j^{th}$ SU and the $i^{th}$ SU, $n_i$ and $n_j$ denote the channels noise, and i corresponds to the cluster head. Equation (13) can be reformulated as

$$Y_i = h_{pi}x_p + \sum_{j=1, j \neq i}^{k} h_{ji}x_j + n \quad (14)$$

where n denotes the total noise in the system. The received signal includes three components: PU signal, other SUs signals, and noise.

### 3.2 Signal Acquisition With Compressive Sensing

The signal acquisition is performed with the compressive sensing to acquire the main information from the original signal and reduce the number of samples for the next processes. We assume that the signals are sparse to meet the sparsity requirement of the compressive sensing. As the processing time and the accuracy of the compressive sensing depend mainly on the sensing matrix selection, we propose to adopt the analog to information converter (AIC) for signal sampling. An analog to information converter is a compressive sensing technique that only samples the signal by multiplying it with a pseudo-random square wave generated from a random number generator. To simply the proposed model, we assume that the signal sampling is performed with the same sensing matrix at each SU node. Acquiring the SU signal consists of multiplying it with a sensing matrix to capture only the main information from the signal. The $i^{th}$ compressed measurements can be expressed as

$$r_i = A_i y_i + e_i \quad i = 1 \dots K \quad (15)$$

where $r_i$ denotes the acquired signal (M, 1) by the $i^{th}$ SU, $A_i$ denotes the sensing matrix (M, N) at the $i^{th}$ SU, $y_i$ denotes the $i^{th}$ SU received signal (N,1), $e_i$ denotes the compression error, and K denotes the number of the cluster heads. Equation (15) can be reformulated when SUs are independent as

$$r_i = A_i h_{pi} x_p + n_i + e_i, i = 1 \dots K \quad (16)$$
$$= B_i x_p + w_i, i = 1 \dots K$$

where $B_i = A_i h_{pi} x_p$, $w_i = n_i + e_i$, and $h_{pi}$ denotes the complex channel gain of the sensing channel between the PU and the i$^{th}$ SU.

With the increase of the number of SUs, the computing complexity increases as well. Some adjacent SUs may have similar characteristics. In order to exploit this feature and decrease the computing complexity, let us organize the SUs in clusters. Each cluster includes a number of SUs and the number of clusters is smaller than the number of SUs. Acquiring the SU signal consists of multiplying signals from the same cluster with a sensing matrix to capture only the main information from the signal. The i$^{th}$ compressed measurements can be expressed as

$$r_i = A_i y_i + e_i, \quad i = 1 \dots K \qquad (17)$$

where $r_i$ denotes the acquired signal (M, 1) by the i$^{th}$ SU, $A_i$ denotes the sensing matrix (M, N) at the i$^{th}$ SU, $y_i$ denotes the i$^{th}$ SU received signal (N, 1), $e_i$ denotes the compression error, and K denotes the number of the cluster heads. Equation (4) can be expressed as

$$\begin{aligned} r_i &= A_i h_{pi} x_p + n_i + e_i, i = 1 \dots K \\ &= B_i x_p + w_i, i = 1 \dots K \end{aligned} \qquad (18)$$

where $B_i = A_i h_{pi} x_p$, $w_i = n_i + e_i$, and $h_{pi}$ denotes the complex channel gain of the sensing channel between the PU and the i$^{th}$ SU.

Therefore, considering cluster heads as nodes, the signal acquisition at each node is performed by a linear projection of the received signal using the AIC assuming independent SUs. As the number of cooperating SUs increases, the system complexity as well as the sensing time increases. To overcome this problem, we propose to discard SUs in faded channels. To do so, let consider L as the total number of cooperating SUs. The signal power of each SU signal is computed before performing the signal sampling and then compared with a threshold. The system is expressed as

$$\begin{aligned} &\text{if } P_i \geq \delta, \text{ keep the SU}_i \text{ signal in the cluster} \\ &\text{if } P_i < \delta, \text{ discard the SU}_i \text{ signal in the cluster} \end{aligned} \qquad (19)$$

where $P_i$ denotes the SU$_i$ signal power, $\delta$ denotes the threshold, and i=1…L. By discarding the SUs with weak signal power, the number of cooperating SUs is reduced from L to J, where J≪ L. In addition, after performing the compressive sensing, the number of samples considered for the sensing is reduced from N to M, where M ≪ N, N is the number of samples of the original signal and M is the measurements number. Thus, the signal power after signal sampling and SUs discarding (MxJxP) is reduced compared to the signal power before sampling and SUs discarding (NxJxP).

### 3.3 Signal Recovery and Spectrum Detection

The compressed signal is considered as the input of the signal detection process. In order to reduce the processing time, recovering the signal and sensing the spectrum are performed simultaneously. For the local spectrum detection, we opt for the power spectral density-based sensing technique [37]. It consists of recovering the power spectral density (PSD), also called power spectrum, from the measurements instead of recovering the whole signal. The information required to decide about the spectrum utilization can be found from the PSD. The power spectral density of a signal, x, is the Fourier transform of the autocorrelation of this signal and it is given by

$$P_r(f) = \int_{-\infty}^{+\infty} E[r(t)r(t-\tau)]e^{-2\pi jf\tau} d\tau \qquad (20)$$

where $P_r$ denotes the power spectrum of the signal, r represents the compressed signal, and f denotes the frequency.

The power spectral density is computed at each node and compared with a predefined threshold to decide about the spectrum occupancy. The system model is expressed as

$$\begin{aligned} &\text{if } P_r \geq \gamma, \text{ PU present} \\ &\text{if } P_r < \gamma, \text{ PU absent} \end{aligned} \qquad (21)$$

where $\gamma$ denotes the detection threshold and $P_r$ denotes the power spectrum. When the power spectrum is higher than the threshold, the channel is considered occupied otherwise it is considered free, and the SUs can transmit their data using this channel. Then, each SU performs its local decision, $d_i$, individually based on the results of the comparison.

## 3.4 Reporting Sensing Results

After performing the spectrum sensing and obtaining the sensing results, these results are forwarded to the fusion center for local decisions. The reporting is performed by sending the local decisions simultaneously from the cluster heads to the fusion center. The reporting mechanism consists of forwarding the individual detection results, $d_i$, to the fusion center [38]. Also, SUs with weak signals are discarded as previously mentioned, which reduces the number of reporting SUs. As a result, the time required by the reporting, called reporting time, is reduced when the number of the reporting heads decreases. After receiving the local decisions, the fusion center uses hard or soft combination rules. Here, we opted for the majority rule, which is based on a voting system to consider the votes with more than the half [39-41].

## 4 RESULTS AND DISCUSSION

The two approaches, cooperative spectrum sensing based compressive sensing and cooperative spectrum sensing without compressive sensing, were implemented and their efficiencies were compared. For the assumptions, we assume the sparsity of the received signals, one PU and several SUs, independent SUs, and one sampling matrix is used. For the performance evaluation, several metrics were used, namely probability of detection, compression ratio, probability of error, and processing time. The probability of detection, $P_d$, is simulated and computed as the ratio of the total number of times the channel is detected as occupied, $N_d$, over the total number of experiments, $N_t$. It is given as

$$P_d = \frac{N_d}{N_t} \quad (22)$$

As previously mentioned, there are three evaluation metrics that can be used to evaluate the efficiency of the detection: probability of detection, probability of false alarm, and probability of miss detection. The probability of error, $P_e$, is the number of errors in detecting correctly the state of a channel. It is expressed as the sum of the errors due to the false alarm detections and miss detections [42]. It is given as

$$P_e = P_{fa} + P_m \quad (23)$$

Examples of the simulation results are presented in Figure 7 to Figure 12. Figure 7 represents the probability of detection, $P_d$, as a function of the number of cooperating SUs for the two approaches: cooperative spectrum sensing based compressive sensing and cooperative spectrum sensing without compressive sensing. As one can see, the probability of detection increases with the increase of the number of cooperating SUs for both approaches, with and without compressive sensing. As expected, the probability of detection is lower for the compressive sensing technique compared to the probability of detection of the technique without compressive sensing. However, the two probabilities become almost equal for a number of SUs greater than 20.

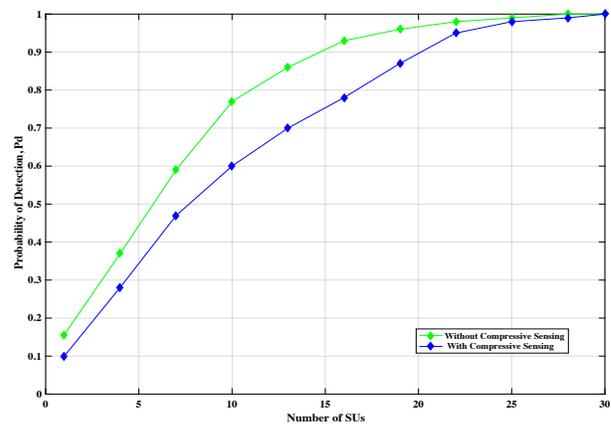

**Figure 7.** Probability of detection against the number of cooperating SUs with and without compressive sensing.

Figure 8 represents the probability of detection corresponding to our model as a function of SNR for different number of SUs. SNR varies from -20 dB to 20 dB, N=1000, M=500, and the compression rate is 50%. As expected, $P_d$ increases with the increase of SNR. For SNR values lower than -5 dB, $P_d$ is under 20% for 5 SUs as well as for 10 SUs. Starting SNR =5 dB, $P_d$ increases achieving higher values for high SNR values. In addition, $P_d$ with 10 SUs is higher than $P_d$ with 5 SUs. Both probabilities become very close at high SNR (SNR $\geq$ 17 dB). Thus, as the number of cooperating SUs increases, the detection rate gets higher and the detection performance gets better.

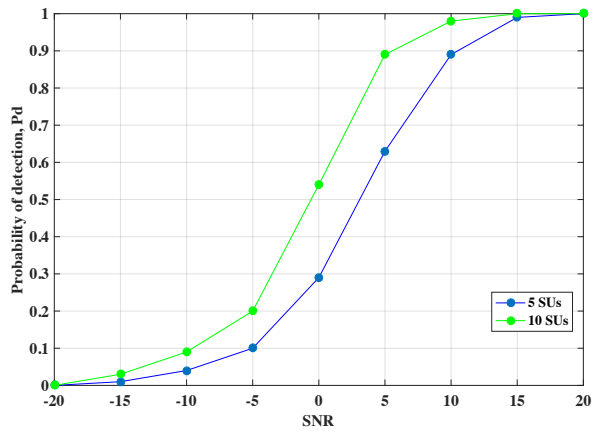

**Figure 8.** Probability of detection against SNR for different cooperating SUs number with compressive sensing.

Figure 9 represents the probability of detection of our model as a function of SNR with 10 SUs for several values of the compression ratio, namely: 10%, 20%, 50%, and 70%. As expected, $P_d$ increases with the increase of SNR. In addition, this probability increases with the decrease of the compression ratio. $P_d$ achieves its maximum value at SNR=2 dB with a low compression ratio of 10% while it achieves that maximum value at SNR=8 dB with a high compression ratio of 70%. Thus, $P_d$ is higher for low compression ratios. Moreover, the probability of detection decreases with compressive sensing, which is due to the fact that detecting with compressive sensing involves a few measurements while detecting without compressive sensing performs on all the samples. For lower SNR, the number of measurements required can be set to a specific value so that it gives similar detection probability. This number of measurements number can be then reduced to achieve that probability of detection while minimizing the required processing time.

Figure 10 represents the probability of error as a function of SNR for different values of the cooperating SUs, which are computed using Equation (26). As can be seen, The probability of error decreases with the increase of SNR. In addition, the probability of error decreases with the increase of the number of cooperating users in the clusters. It is low for 10 SUs and high for 5 SUs at high SNR, which can be explained by the fact that more cooperating users sensing the spectrum results in less error detection rates. Thus, cooperative spectrum sensing based clustering scheme reduces the error rates in terms of detecting unused spectrum with large number of users.

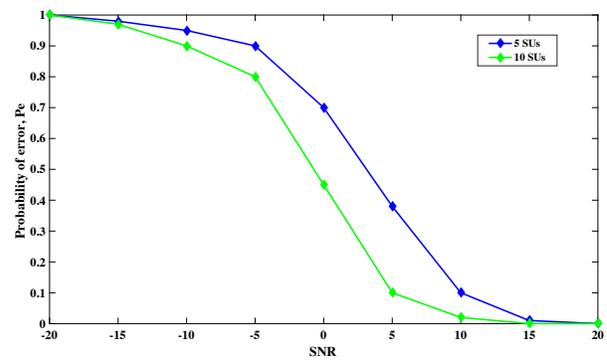

**Figure 10.** Probability of error as a function of SNR for cooperating SUs number with compressive sensing.

Figure 11 represents the probability of error of our model as a function of SNR with 10 SUs for several values of the compression ratio, namely: 10%, 20%, 50%, and 70%. As expected, $P_e$ decreases with the increase of SNR. In addition, the probability of error decreases with the decrease of the compression ratio. It achieves its minimum value at SNR=2 dB with 10% as a compression ratio while it achieves that minimum value at SNR=8 dB with 70% of compression ratio. Thus, our proposed approach improves the accuracy of the detection by reducing the probability of error.

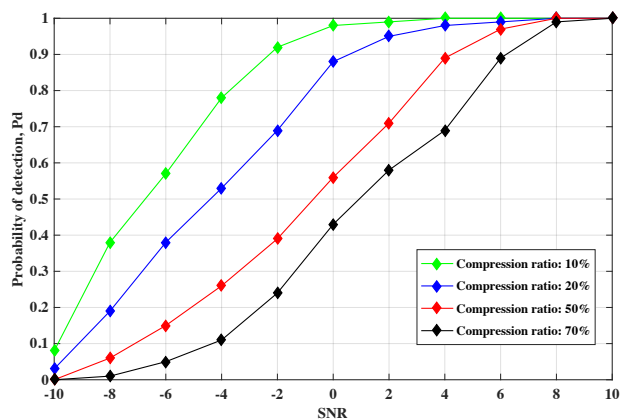

**Figure 9.** Probability of detection against SNR for different compression ratio (M/N).

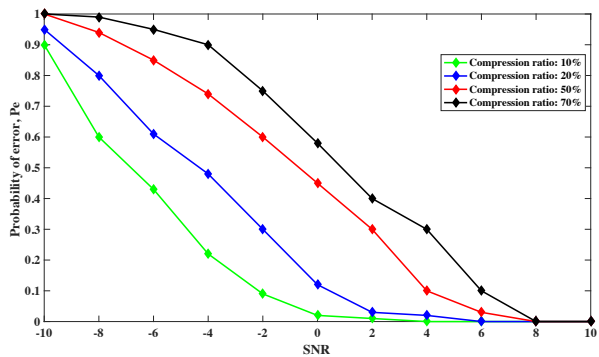

**Figure 11.** Probability of error as a function of SNR for different compression ratio.

Figure 12 represents the processing time as a function of the number of cooperating SUs for different values of the compression ratio, namely: 1%, 10%, and 50%. As expected, the processing time increases with the increase of the number of SUs. This figure also shows that the cooperative spectrum sensing without compressive sensing requires a great deal of processing time compared to the cooperative spectrum sensing with compressive sensing. For instance, for 20 SUs, cooperative spectrum sensing with low compression ratio (10%) requires an average of 4 ms, but the cooperative spectrum sensing with a compression ratio of 50% requires 8 ms, which is 2 times higher. In addition, the processing time required by the cooperative spectrum sensing without compressive sensing (M/N=100%) to perform is about 14 ms, which is 4 times slower than sensing with low compression ratio. Thus, our proposed approach reduces the processing time required by the spectrum sensing.

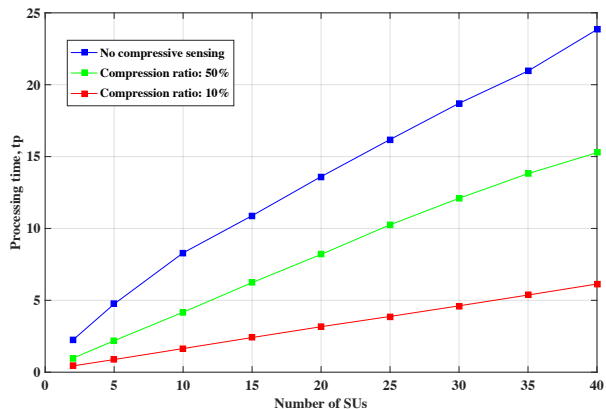

**Figure 12.** Processing time as a function of the number of SUs for different compression ratio (M/N).

To sum up, Table 1 summarizes the results of the simulation.

**Table 1.** Performance evaluation of the proposed scheme

|  | Low compression ratio M/N=10% | High compression ratio M/N =70% | Sensing without compressive sensing |
|---|---|---|---|
| Probability of detection (%) SNR=0 dB #SUs = 10 | High (97%) | Medium (43%) | High (99%) |
| Processing time (ms) SNR=0 dB #SUs = 10 | Fast (2 ms) | Slow (10 ms) | Slow (8 ms) |

These examples of results show that our proposed technique is more efficient and faster in detecting high dimensional sparse signals with even low compression ratio. Through comparing the different methods, spectrum sensing based compressive sensing can be chosen as an efficient method for fast and accurate cooperative spectrum sensing.

## 5 CONCLUSIONS

Cooperative spectrum sensing has been proposed to enhance the sensing performance when a number of secondary users (SUs) are observing the same band of interest. Existing cooperative spectrum sensing techniques are still not efficient in terms of processing time, sampling complexity, and communication overhead. Thus, fast signal acquisition approaches are highly needed to reduce the processing time while achieving high detection. In this paper, we have proposed a cooperative spectrum sensing technique based on compressive sensing. The results of the proposed technique are discussed and compared to those of the cooperative spectrum sensing without involving compressive sensing. Metrics used to assess the efficiency of our technique include the probability of detection, compression ratio, probability of error, and processing time [43]. The simulation results show that sensing cooperatively and compressively the radio spectrum speeds up the processing time while achieving good detection performance with low compression ratio and low error rates. For future works, the proposed model will be enhanced by considering co-headship in the cluster instead of one single head in order to reduce the workload on a single cluster head. This approach aims at reducing the workload on a single cluster head by sharing it between cluster heads. In addition, considering more complicated scenarios with several PUs and SUs using the wideband spectrum in complex and noisy transmission environment to overcome the limitations of this work.